\documentclass[10pt,twocolumn,letterpaper]{article}

\usepackage{cvpr}              

\usepackage{graphicx}
\usepackage{amsmath}
\usepackage{amssymb}
\usepackage{booktabs}
\usepackage[table,x11names]{xcolor}
\usepackage{color}
\usepackage{colortbl}
\usepackage{afterpage}
\usepackage{bbm}

\newif\ifdraft
\drafttrue

\newcommand{\myparagraph}[1]{\smallskip\noindent\textbf{#1}}

\newcommand{\calC}{\mathcal{C}}
\newcommand{\R}{\mathbb{R}}
\DeclareMathOperator{\dist}{dist}

\usepackage[pagebackref,breaklinks,colorlinks]{hyperref}
\usepackage{amsmath}

\usepackage[capitalize]{cleveref}
\crefname{section}{Sec.}{Secs.}
\Crefname{section}{Section}{Sections}
\Crefname{table}{Table}{Tables}
\crefname{table}{Tab.}{Tabs.}

\definecolor{LightGray}{rgb}{0.9,0.9,0.9}
\def\cvprPaperID{8161} 
\def\confName{CVPR}
\def\confYear{2023}

\begin{document}


\title{Topology-Guided Multi-Class Cell Context Generation for Digital Pathology}

\author{Shahira Abousamra$^{1}$, Rajarsi Gupta$^{2}$, 
Tahsin Kurc$^{2}$, 
Dimitris Samaras$^{1}$, Joel Saltz$^{2}$ and Chao Chen$^{2}$ \\
{\tt\small $^1$Stony Brook University, Department of Computer Science, USA} \\
{\tt\small $^2$Stony Brook University, Department of Biomedical Informatics, USA} \\
}

\maketitle

\begin{abstract}
In digital pathology, the spatial context of cells is important for cell classification, cancer diagnosis and prognosis. To model such complex cell context, however, is challenging. Cells form different mixtures, lineages, clusters and holes. To model such structural patterns in a learnable fashion, we introduce several mathematical tools from spatial statistics and topological data analysis. We incorporate such structural descriptors into a deep generative model as both conditional inputs and a differentiable loss. 
This way, we are able to generate high quality multi-class cell layouts for the first time. 
We show that the topology-rich cell layouts can be used for data augmentation and improve the performance of downstream tasks such as cell classification. 


\end{abstract}

\section{Introduction}
\label{sec:intro}

Deep learning has advanced our learning ability in digital pathology. 
Deep-learning-based methods have achieved impressive performance in various tasks including but not limited to: cell detection and classification~\cite{Abousamra_2021_ICCV, Hung:2020:BMC:nucleus-detection:faster-rcnn,Yousefi:2019:ISBI:cell-detect:faster-rcnn, Hofener:2018:cell-detection}, nuclei instance segmentation~\cite{hou-gan-2019-cvpr,Graham:2020:ITMI:nucleus-segmentation,graham:2019:hover,RAZA:2019:MIA:nucleus-segmentation,Kumar:2017:ITMI:nucleus-segmentation,naylor:2018:ipmi:cell-sup-seg,Qu:midl:2019:point_seg,yoo:miccai:2019:point_seg,Tian:miccai:2020:point_seg,chamanzar:isbi:2020:point_seg}, survival prediction and patient outcome ~\cite{survival:Wulczyn:2020:plos,survival:Abbet:2020:miccai,survival:Huidong:2022:bioinformatics,outcome:cooper:2018:pnas}, interpretation of multiplex immunohistochemistry and immunofluorescence imagery~\cite{abousamra:multiplex:2020:isbi,Fassler2020DeepLI,ghahremani2022deepliifui,ghahremani2022deep}  and many others.

Despite the rapid progress in recent years, pathology image analysis is still suffering from limited observations. The available annotated images are still scarce relative to the highly heterogeneous and complex tumor microenvironment driven by numerous biological factors. The limitation in training data constraints a learning algorithm's prediction power. To this end, one solution is to train generative models that can generate realistic pathology images to augment existing data. Generative models have been proposed to help learning methods in various tasks such as nuclei segmentation \cite{hou-gan-2019-cvpr, Sharp-GAN-2022-isbi}, survival prediction \cite{gan:survival:Tomoki} and cancer grade prediction \cite{Wei:cancer-grade:gan}. 

Generating pathology images usually involves two steps: (1) generating spatial layout of cells and (2) filling in stains and textures inside and outside cell nuclei masks. 
Most existing methods only focus on the second step. 
They either generate random cell positions \cite{hou-gan-2019-cvpr} or directly copy nuclei masks from existing images \cite{gong-style-consistent-2021-wacv}. These methods miss the opportunity to learn the rich cell spatial context carrying critical information about cancer biology.

\begin{figure}[t]
  \centering
  \centerline{\includegraphics[width=1\linewidth]{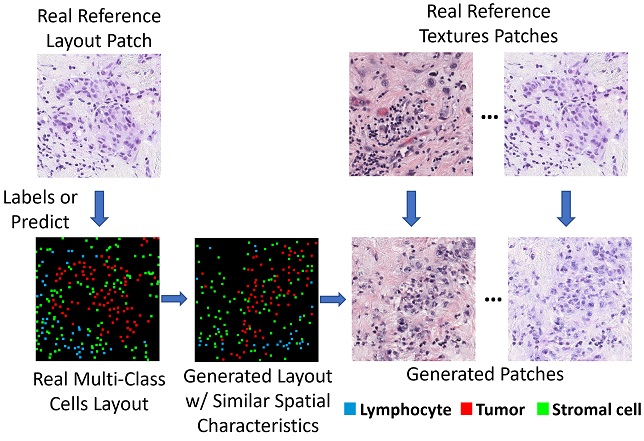}}
\caption{Overview of our multi-class cell context generator. }
\label{fig:teaser}
\end{figure}

\begin{figure*}[t]
  \centering
  \centerline{\includegraphics[width=1.0\linewidth]{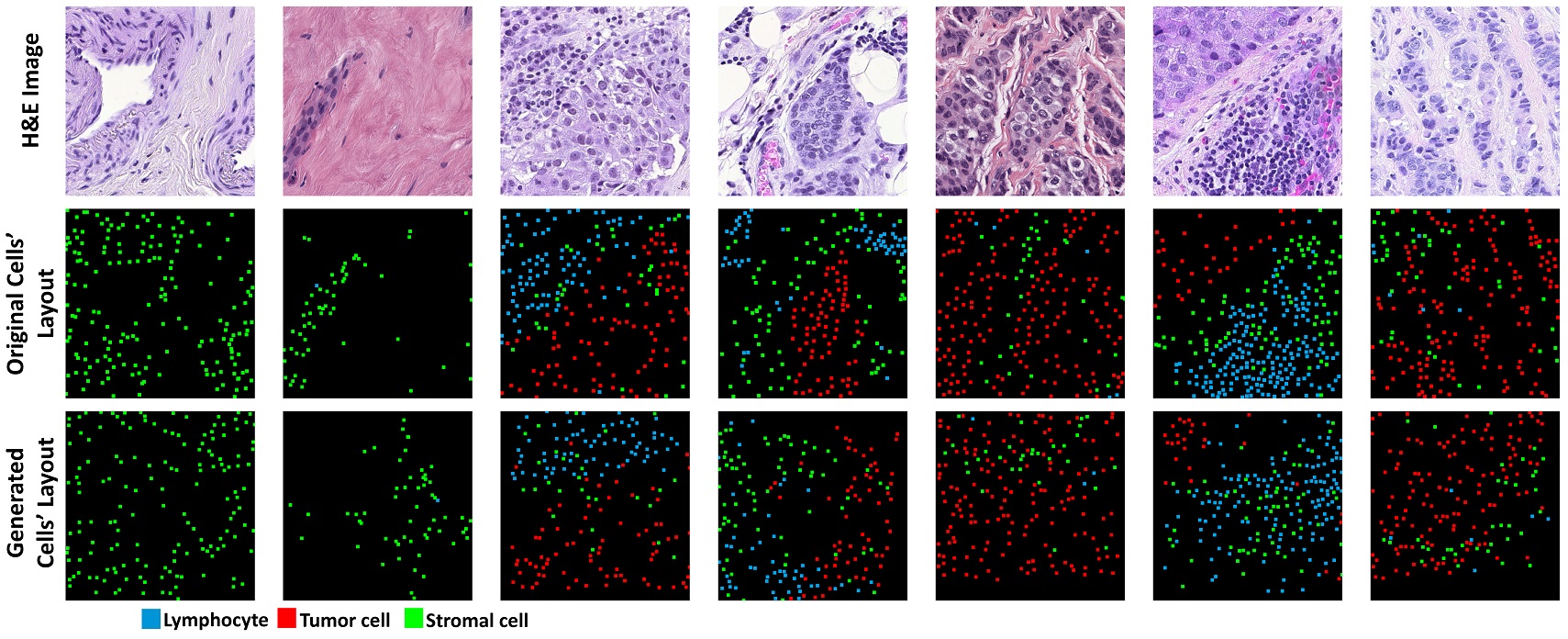}}
\caption{Sample results from our cell layout generator. Our generated samples have similar spatial characteristics as the corresponding reference layouts.}
\label{fig:results-layout}
\end{figure*}

Spatial context includes how different types of cells (tumor, lymphocyte, stromal, etc)  are distributed around each other, as well as how they form different structural patterns such as clusters, holes and lineages. Plenty of evidence have demonstrated the importance of spatial context in cancer diagnosis and prognosis \cite{nawaz2016computational,Yuan:2012:cell-classify}. One good example is the clinical significance of tumor infiltrating lymphocytes (TILs), i.e., lymphocytes residing within the border of invasive tumors \cite{salgado2015evaluation,saltz2018spatial,Shibutani:2018:imp-classify:TIL-outcome,Stanton:2016:imp-classify:TIL-outcome}. 
The spatial distribution of stromal cells in the vicinity of tumor has been shown to be directly related to cancer outcomes \cite{Yuan:2012:cell-classify,Rogojanu:2015:imp-classify}. 
Tumor budding, i.e., the presence of isolated or small clusters of tumor cells at the invasive tumor front, is a prognosis biomarker associated with an increased risk of lymph node metastasis in colorectal carcinoma and other solid malignancies \cite{lugli2020tumour}. In prostate cancer tissue samples, plenty of loopy cellular structures are formed corresponding to glands. Their integrity, known as the Gleason score, is a good indicator of cancer progression \cite{WRIGHT20092702:gleasonscore}. 


Given the biological significance of cell spatial context, we hypothesize that being able to model and generate cell configurations will benefit various downstream tasks. To model the complex cell spatial context, the main challenge is the limited information one can rely on --coordinates and types of cells. This makes it hard for even powerful deep learning methods \cite{li2021spgan} to learn the underlying distribution.
To better model the spatial context, we argue that principled mathematical machinery has to be incorporated into the deep learning framework. Formally, we introduce the classic K-function from spatial statistics \cite{Baddeley:2015:book:spatial-point-pattern}, as well as the theory of persistent homology \cite{edelsbrunner2010computational}, to model the spatial distribution of multi-class cells and their structural patterns. These mathematical constructs have been shown to correlate with clinical outcomes \cite{aukerman2021persistent}. However, they have not been used in the generation of pathology images.

We incorporate these spatial topological descriptors into a deep generative model. Our generative model takes an input pathology image and generates a new cell layout with similar spatial and topological characteristics. To enforce the expected spatial characteristics, we propose a novel \emph{cell configuration loss} based on the persistent homology and spatial statistics of input cell spatial configuration. The loss compares the generated and the reference cell configurations and match their topology in view of a topological measure called persistence diagram. The loss enforces holes in the generated cell configuration to be one-to-one matched to holes in the reference cell configuration, i.e., having similar shapes and density. 

A direct topological matching via persistence diagrams is agnostic of the cell type composition. This is undesirable; we do not want to match a tumor cell hole to a stromal cell hole. To this end, we also incorporate spatial statistics measure, i.e., cross K-functions, into the loss. This way, holes composed of different types of cells are matched properly. 
Using the generated cell spatial configuration, we generate the nuclei mask, staining and texture. 

See \cref{fig:teaser} for an illustration of the generation pipeline. Also see \cref{fig:results-layout} for examples of the generated cell layouts. The generated cell layouts have very similar spatial and structural characteristics as the reference/input image. This is not guaranteed with previous methods using randomly generated masks. In the experiment section, we provide comprehensive comparisons to verify the benefit of our method. We will also show that the augmented images can be used to train downstream tasks such as cell classification. 

\begin{figure*}[t]
  \centering
  \centerline{\includegraphics[width=1\linewidth]{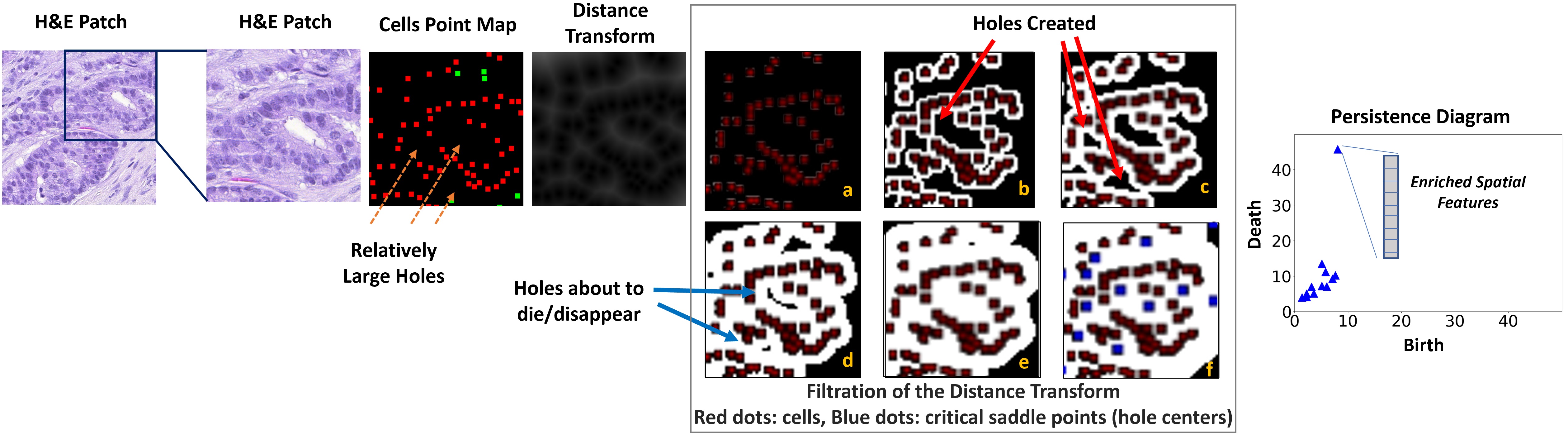}}
\caption{Illustration of the filtration process of the distance transform map to obtain the persistence homology. Red dots are the tumor cells in the original image. The blue dots in the last figure (f) are the centers for the holes, the saddle points which are obtained once a hole dies or disappears.}
\label{fig:filtration}
\end{figure*}

To summarize, our contributions are as follows:
\begin{itemize}
    \item We propose the first generative model to learn cell spatial context from pathology images. 
    \item We introduce multi-class spatial context descriptors based on spatial statistics and topology. These descriptors are used as conditional input for the generator.
    \item We propose a novel cell configuration loss function to enforce the desired behavior of spatial distribution and topology. The loss matches holes of generated cell layout and holes of the reference cell layout, in shape, density, and  cell type composition.
    \item We show that the generated layouts can be used to generate synthetic H\&E images for data augmentation.
    We show the efficacy of the augmentation data in downstream tasks such as cell classification.
\end{itemize}

We stress that the benefit of modeling cell spatial context is beyond data augmentation. Modeling the spatial context will provide the foundation for better understanding and quantifying the heterogeneous tumor microenvironment, and correlate with genomics and clinical outcomes. This work is one step towards such direction.


\section{Related Work}
\label{sec:relatedwork}
Generative models have been broadly used in medical imaging. Within the context of digital pathology, different methods~\cite{hou-gan-2019-cvpr, tsirikoglou-inter-organ-miccai-21,Sharp-GAN-2022-isbi} have been proposed to use generated images as an augmentation for nuclei or tissue structure segmentation. Most of these methods, however, overlook generating spatial configuration of cells.
Several methods~\cite{hou-gan-2019-cvpr,renal-cspa-2020, Sharp-GAN-2022-isbi} creates randomly distributed nuclei masks before generating staining and texture.
When the downstream task is not nuclei segmentation, one may generate randomly distributed masks of other structures, e.g., glands~\cite{deshpande-safron-mia-22}. 
Another category of methods generates new images using nuclei masks from reference images and only synthesize the staining and textures~\cite{tsirikoglou-inter-organ-miccai-21, sian-wang-corr-22, boyd-region-guided-cyclegan-miccai-22}.
Gong et al.~\cite{gong-style-consistent-2021-wacv} randomly deform the nuclei mask from a reference image. These methods, however, still use the same cell positions. 
All these methods either use the original cell spatial configuration or generate random configurations.
To the best of our knowledge, \emph{our method is the first to learn to generate the cell spatial and structural configurations.} 


\myparagraph{Topology-aware losses.}
Persistent-homology-based losses have been used to enforce topological constraints in image segmentation and generation \cite{hu2019,hu2021,wang2020topogan}. These methods focus on thin structures like vessels and road networks. They are not applicable to modeling structural patterns in cells configuration. Perhaps the closest work to ours is TopoGAN~\cite{wang2020topogan}. It learns to generate thin structures whose numbers of holes/loops match those of the real images. The key difference of our method is that our topological features are adapted to the cell layout setting. The persistence diagram is enriched with cell density and multi-class cell composition information. The loss is also based on the enriched persistence diagrams. This is critical to our success. 


\section{Method}
\label{sec:method}
\begin{figure}[t]
  \centering
  \centerline{\includegraphics[width=1\linewidth]{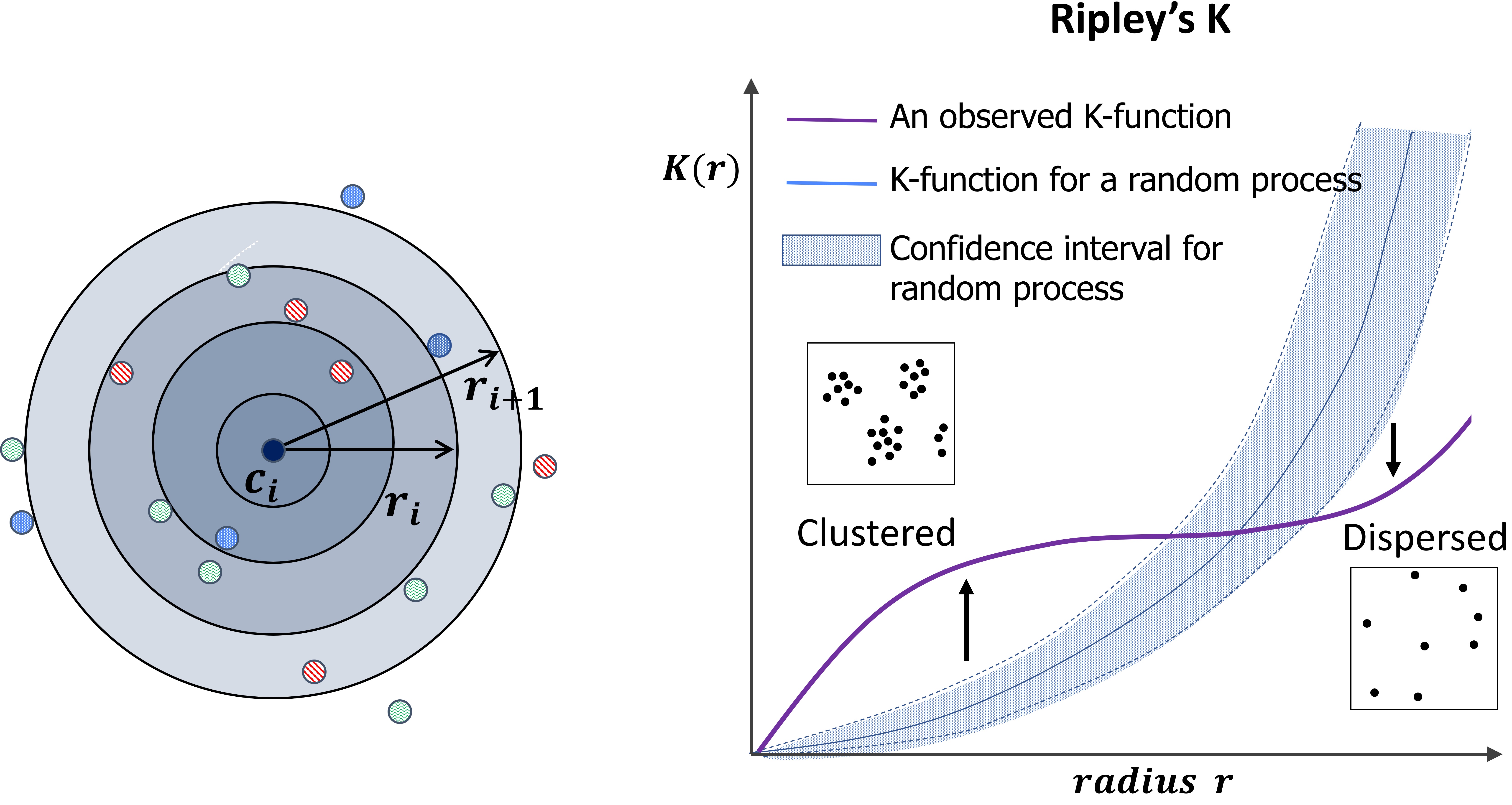}}
\caption{Ripley’s K function. The K-function considers the number of neighboring target points (cells) of different classes at increasing radii from a source. The K-function can indicate the spatial distribution clustering or scattering. Left: an illustration of the computation of K-function at radius $r_i$. Right: An observed K-function (purple) and a K-function of a Poisson random point process (blue). The distribution is clustered/scattered when the observed K-function is higher/lower than the K-function of the random process, respectively.}
\label{fig:k-func}
\end{figure}

\begin{figure*}[t]
  \centering
  \centerline{\includegraphics[width=0.9\linewidth]{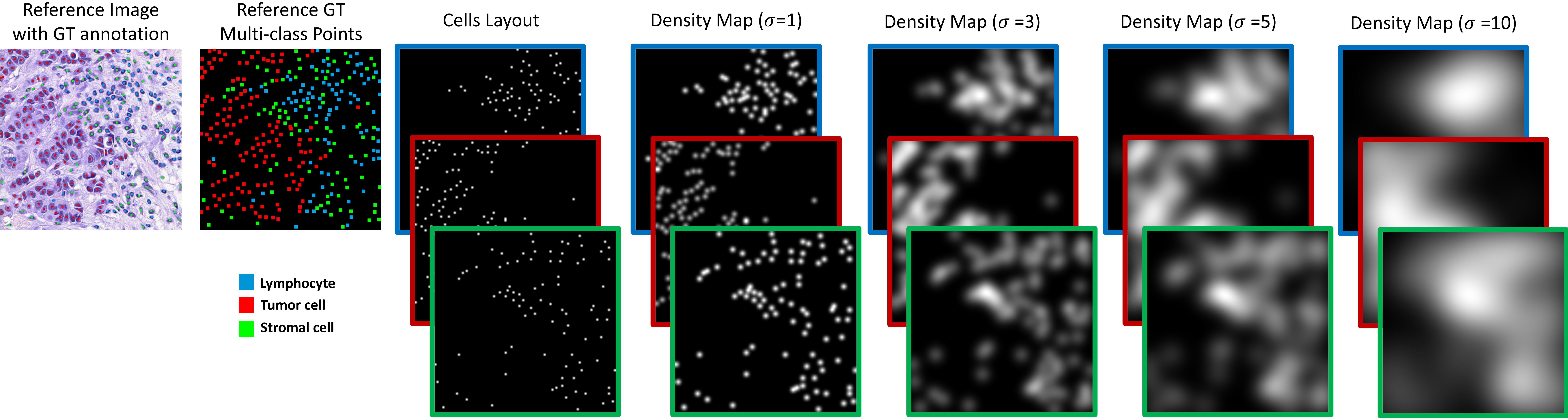}}
\caption{Illustration of the multi-scale density maps generated from a dot map of cells with different Gaussian kernel standard deviations.}
\label{fig:dmap-all}
\end{figure*}

Assume we are given a \emph{cell layout}, i.e., a set of multi-class cells distributed in the image domain. The spatial configuration of these cells includes their structural organization, as well as the spatial distribution of different cell classes. 
Given a reference layout, our method generates a multi-class cell layout with a similar configuration. The generated layout can be used for different purposes including data augmentation. Our model takes as input not only the reference layout, but also a set of spatial descriptors collected from the reference layout. 
Furthermore, for the training of the generator, we propose a loss function to match topological features in the generated and reference layouts. Minimizing such loss will ensure the generated layout has similar structural patterns as the reference layout.


\begin{figure*}[t]
  \centering
  \centerline{\includegraphics[width=1.0\linewidth]{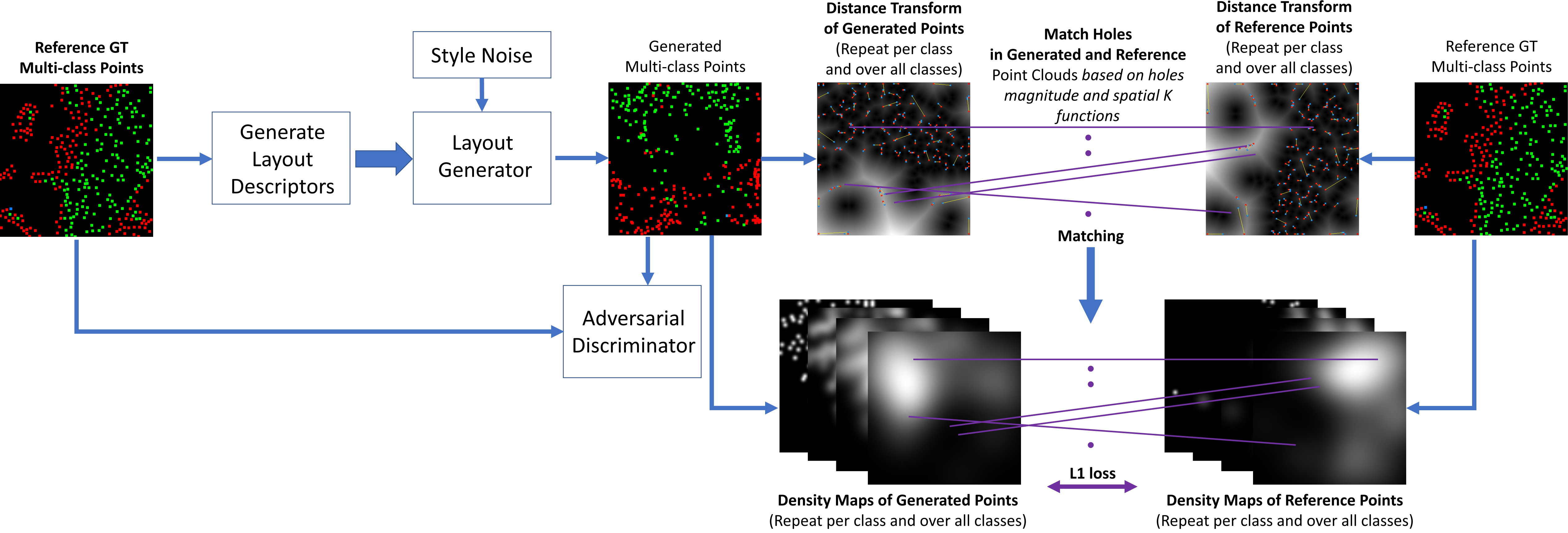}}
\caption{Training of our Cell Layout Generator.}
\label{fig:layout-train}
\end{figure*}

In Section \ref{sec:method:descriptors}, we will introduce different spatial descriptors we will use, based on the theory of persistent homology and the classic spatial statistics. In Section \ref{sec:method:generator}, we will introduce the proposed neural network generator, as well as how these spatial descriptors are incorporated to ensure that the generated layout has a desired configuration.

\subsection{Cell Configuration Descriptors}
\label{sec:method:descriptors}
Our configuration descriptors should capture (1) structural patterns such as clusters and holes of a reference cell layout; and (2) how different types of cells are distributed with regard to other types. 
These structural and multi-class distribution is part of what pathologists study when they inspect a histology image. We formalize such information into two descriptors: cross K-function features and enriched persistence diagram features.

\myparagraph{Spatial statistics features: cross K-functions of cells.}
We first characterize the relative distribution across different classes of cells, e.g., how close are lymphocytes distributed surrounding tumor cells. We use the cross K-function from the classic spatial statistics \cite{Baddeley:2015:book:spatial-point-pattern}. See~\cref{fig:k-func}. Given two cell classes (source class and target class), the cross K-function measures the expected number of neighboring target class cells within different radii of a source class cell. Formally, denote by $\calC_s$ and $\calC_t$ the set of source cells and the set of target cells, respectively. The cross K-function at radius $r$ is defined as:
\begin{equation}
  K_s^t(r) = A \sum_{c_s\in \calC_s}\sum_{c_t\in \calC_t} \mathbbm{1}\{\dist(c_s,c_t)<r\}
  \label{eq:important}
\end{equation}
where $A$ is a normalization term depending on the image area and the sizes of $\calC_s$ and $\calC_t$. $\mathbbm{1}\{\cdot\}$ is the indicator function.\footnote{We simplify the definition of K-function by ignoring the edge correction term.} 
The cross K-function is computed for each pair of classes. Note that when the source and target represent the same class, the K-function is measuring how much a particular class of cells is clustered.
In practice, we vectorize the K-function by sampling at a finite set of radii. We note that K-function has previously been used in cell classification task \cite{Abousamra_2021_ICCV}, but it has not been used for cell configuration characterization and cell layout generation.

We also use a location-specific K-function,
\begin{equation}
  K^t(r,x) = A' \sum_{c_t\in \calC_t} \mathbbm{1}\{\dist(x,c_t)<r\}.
  \label{eq:cell-k-function}
\end{equation}
It describes the distribution of target class cells surrounding a specific location $x$. This will be used for the characterization of  holes identified by persistent homology.

\myparagraph{Topological features: enriched cell  persistence diagrams.}
We propose topological features characterizing gaps and holes distributed in a cell layout. These topological structures provide unique structural characterization of the cell layout, as evident in sample cell layouts (see \cref{fig:results-layout} second row). We use the theory of persistent homology which captures holes and gaps of various scales in a robust manner. To adapt to cell configuration characterization, we propose to enrich the output of persistent homology with spatial distribution information, so that the topological structures are better characterized. 

We briefly introduce persistent homology in the context of cell layout characterization. Please refer to \cite{edelsbrunner2010computational} for more details. Given a cell layout $\calC$ with holes in it, we first compute a distance transform from the cells, $f(x) = \min_{c\in \calC} \dist(x,c)$ (see \cref{fig:filtration}). Holes essentially correspond to salient local maxima of the distance transform. To capture these salient holes, we threshold the image domain $\Omega\subset \R^2$ with a progressively increasing threshold $t$. As the threshold increases, the thresholded domain $\Omega^t = \{x\in \Omega\mid f(x)\leq t\}$ monotonically grows from empty to the whole image domain. It essentially simulates the progress of growing disks centered at all cells with an increasing radius $t$. Through the process, different holes will appear (be born) and eventually are sealed up (die). Persistent homology captures all these holes and encodes their information into a 2D point set called a \emph{persistence diagram}. Each hole in the cell layout corresponds to a 2D point within the diagram, whose coordinates are the birth and death times (thresholds) of the hole. Holes with long life spans are considered more salient. See \cref{fig:filtration} for an illustration of the filtration and the corresponding persistence diagram.
Note we only focus on 1D topology, i.e., holes. Clusters of cells can also be described with 0D topology (connected components in the growing $\Omega^t$). We do not think 0D topology is necessary as the spatial statistics feature implicitly characterizes the cell cluster structures.

While persistent homology captures all possible holes in a cell layout, the persistence diagram alone does not really describe the holes in full details. Intuitively, the birth and death times of a point in the diagram only measure the compactness along the boundary and the size of the hole. We propose to enrich the diagram with additional information regarding density and spatial statistics. In particular, for each hole, we focus on its corresponding local maximum of the distance transform. Note this local maximum is the location at which the hole disappears (dies), and its function value is the death time of the hole. It roughly represents the center of the hole. We compute the location-specific K-function (\cref{eq:cell-k-function}) for the local maximum. It essentially characterizes cell class composition surrounding the hole. Furthermore, we compute the multi-scale cell density function at the local maximum, namely, the cell kernel density function estimated with different bandwidths. As shown in \cref{fig:dmap-all}, these multi-scale density functions characterize cell distribution at different scales regarding the hole of interest.

By attaching the spatial statistic and multi-scale density with each persistent point, we compute an \emph{enriched cell persistence diagram}. See \cref{fig:filtration} for an illustration. This diagram will be used in our cell configuration loss. Details will be introduced in \cref{sec:method:generator}. 

\subsection{Deep Cell Layout Generator}
\label{sec:method:generator}
Next, we introduce our deep cell layout generator. The framework is illustrated in \cref{fig:layout-train}. 
From a reference cell layout, we extract spatial descriptors, including persistence diagrams and spatial statistics. We compute different diagrams for different cell classes separately. These diagrams are all used. 
The  generator takes in vectorized spatial descriptors and style noise, and outputs the coordinates of points in the generated layout. 
To vectorize a persistence diagram, we convert it into a histogram with predefined buckets of persistence range values. This takes care of the variation in persistence diagram size across different  point sets. Since larger 1D topological features (i.e. holes) usually have smaller frequency compared to smaller ones, we use the log of the histogram to account for the tail effect. 

The generator backbone model is a modified version of a state-of-the-art point cloud generative model called SP-GAN~\cite{li2021spgan}. SP-GAN is trained with global and local priors. The global prior is the initial input point coordinates which are a fixed set of points sampled from a unit sphere. The local prior is a latent encoding that determines the style. The generator architecture consists of a set of graph attention modules that act on the global prior or points coordinates, intervened with adaptive instance normalization blocks using the local prior or the style embedding. The final generated point coordinates are the output of an MLP block.
Note that our method is agnostic to the backbone. In principle, we can use any other conditional generative model.

Next we introduce a novel loss function that enforces the generated layout to have a matching configuration with the reference layout.

\begin{figure*}[t]
  \centering
  \centerline{\includegraphics[width=1.0\linewidth]{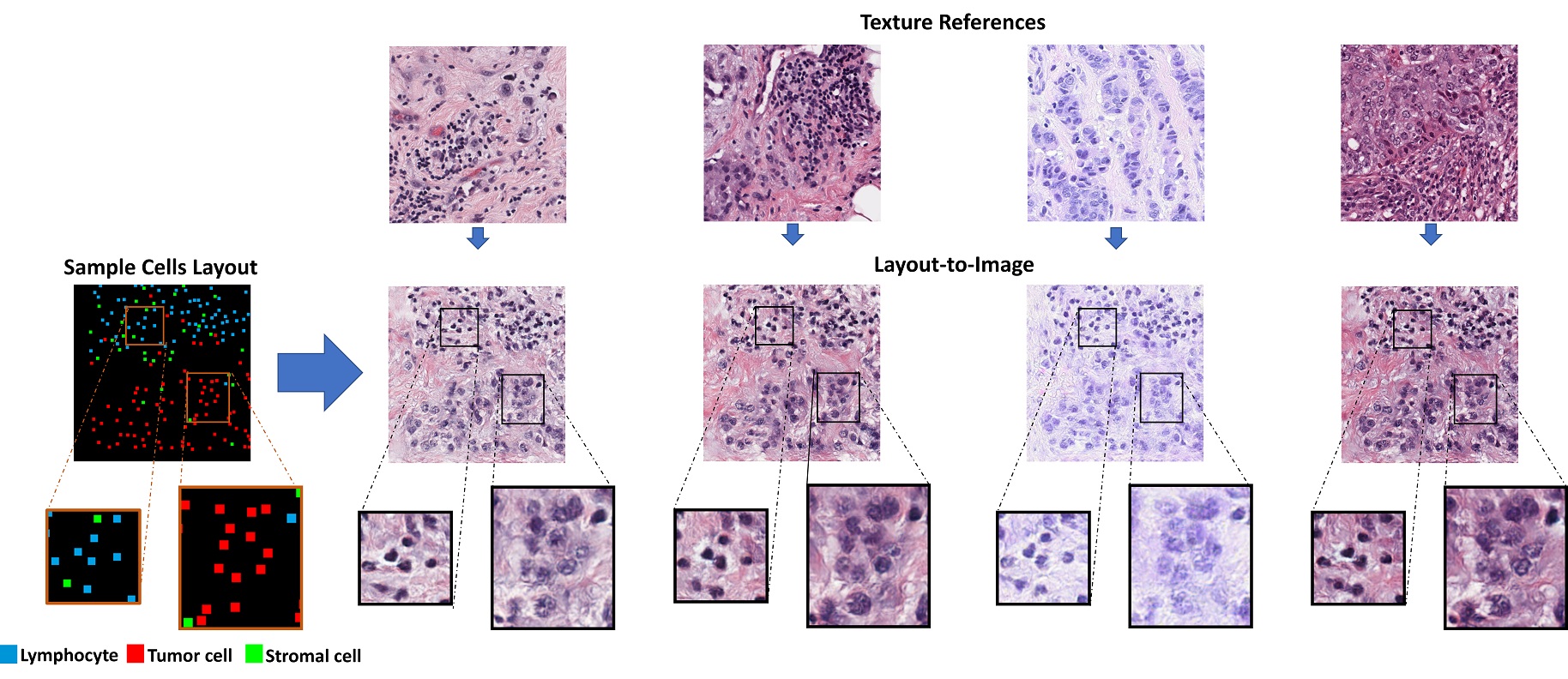}}
\caption{Sample Results from Cells' Layout-to-Image Generator}
\label{fig:results-texture}
\end{figure*}

\paragraph{The cell configuration loss.}
We define the cell configuration loss as the matching distance between the enriched cell persistence diagrams of the generated layout and the reference layout. This is an extension of the classic Wasserstein distance between persistence diagrams \cite{cohen2005stability,cohen2010lipschitz}.

Recall in an enriched diagram, each point (representing a hole) has not only birth/death times, but also additional attributes including location-specific cross K-function and multi-scale density function. We use K-function to match generated holes and reference holes. Then we use the density function between matched holes as a loss to control the generated points. This design choice is well justified; K-function helps identify holes with matching contexts. Meanwhile, for matched holes, using multi-scale density functions as the loss can more efficiently push generated points, thus improving the generator.

In particular, we compute the persistence diagrams $Dgm_{gen}$ and $Dgm_{ref}$ from the generated and reference layouts, respectively. 
Next, we find an optimal matching between the two diagrams. Assume the two diagrams have the same cardinality. We compute 
\begin{equation}
\gamma^\ast = \arg\min_{\gamma\in \Gamma} \sum\nolimits_{p\in Dgm_{gen}} \dist_K(p, \gamma(p)), 
\label{eq:opt-matching}
\end{equation}
where $\Gamma$ is the set of all one-to-one mapping between the two diagrams. The distance $\dist_K(p,\gamma(p))$ is the Euclidean distance between the K-function vectors of the two holes represented by the persistence points $p\in Dgm_{gen}$ and $\gamma(p)\in Dgm_{ref}$. In other words, we find an optimal matching between the diagrams using the K-function distance between holes. If the two diagrams have different cardinalities, the unmatched holes will be matched to a dummy hole with zero persistence. 

Once the optimal matching is found, we define the configuration loss as 
\begin{equation}
L_{CC} = \sum\nolimits_{p\in Dgm_{gen}} \dist_{den}(p,\gamma^\ast(p)),
\label{eq:spatial-layout-loss}
\end{equation}
in which $\dist_{den}(p,\gamma^\ast(p))$ is the distance between the multi-scale density of the two matched holes. 


During training, 
for each pair of generated and reference layouts, we compute their enriched diagrams and find the optimal matching $\gamma^\ast$ using Hungarian method. Then we optimize the loss in Eq.~\ref{eq:spatial-layout-loss}. This essentially moves points in the generated layout so that each hole has a similar multi-scale density as its matched hole. \cref{fig:layout-train} illustrates the loss. 

\section{Experiments}
\label{sec:experiments}
\setlength{\tabcolsep}{4pt}
\begin{table*}
\small
\begin{center}
\begin{tabular}{|p{0.2\linewidth}|p{0.045\linewidth}|p{0.045\linewidth}|p{0.045\linewidth}|p{0.045\linewidth}|p{0.045\linewidth}|p{0.045\linewidth}|p{0.045\linewidth}|p{0.045\linewidth}|}
\hline

& \multicolumn{4}{p{0.22\linewidth}|}{\centering {PD - EMD $\downarrow$}}  
& \multicolumn{4}{p{0.22\linewidth}|}{\centering {PD - CCMD $\downarrow$}}  
\\
\hline
 Method  & \centering{Infl.} & \centering{Epi.} & \centering{Stro.} & \cellcolor{LightGray}\centering{Mean}   
 & \centering{Infl.} & \centering{Epi.} & \centering{Stro.} & \cellcolor{LightGray}\centering{Mean}   
 \cr
\hline
\raggedright{w/o Spatial Descriptors +w/o Matching Loss}
& \centering{0.28} & \centering{{0.082}} & \centering{0.19} & \cellcolor{LightGray}\centering{\textbf{0.184}} 
& \centering{0.80} & \centering{1.74} & \centering{{1.66}} & \cellcolor{LightGray}\centering{1.4} 
\cr
w/o Matching Loss  
& \centering{0.249} & \centering{0.203} & \centering{0.156} & \cellcolor{LightGray}\centering{0.202} 
& \centering{{0.90}} & \centering{{1.69}} & \centering{{1.79}} & \cellcolor{LightGray}\centering{{1.46}} 
\cr
w/o K-function Descriptor 
& \centering{{0.237}} & \centering{0.167} & \centering{0.17} & \cellcolor{LightGray}\centering{0.191} 
& \centering{{0.75}} & \centering{{1.74}} & \centering{{1.77}} & \cellcolor{LightGray}\centering{{1.42}} 
\cr
Ours 
& \centering{0.246} & \centering{0.141} & \centering{{0.165}} & \cellcolor{LightGray}\centering{\textbf{0.184}} 
& \centering{{{0.74}}} & \centering{{{1.64}}} & \centering{1.71} & \cellcolor{LightGray}\centering{{\textbf{1.36}}} 
\cr
\hline
\end{tabular}
\caption{Evaluation of persistence diagram of generated cell layout compared to reference layouts in BRCA-M2C dataset.}
\label{table:layout-eval-PD}
\end{center}
\end{table*}

\setlength{\tabcolsep}{4pt}
\begin{table*}
\small
\begin{center}
\begin{tabular}{|p{0.20\linewidth}|p{0.045\linewidth}|p{0.045\linewidth}|p{0.045\linewidth}|p{0.045\linewidth}|p{0.045\linewidth}|p{0.045\linewidth}|p{0.045\linewidth}|p{0.045\linewidth}|p{0.045\linewidth}|p{0.045\linewidth}|p{0.045\linewidth}|p{0.045\linewidth}|}
\hline

& \multicolumn{4}{p{0.22\linewidth}|}{\centering {Cross K-function MAE $\downarrow$}}  
& \multicolumn{4}{p{0.22\linewidth}|}{\centering {Cross K-function RMSE $\downarrow$ }}  
\\
\hline
 Method  & \centering{Infl.} & \centering{Epi.} & \centering{Stro.} & \cellcolor{LightGray}\centering{Mean}   
& \centering{Infl.} & \centering{Epi.} & \centering{Stro.} & \cellcolor{LightGray}\centering{Mean}   
 \cr
\hline
\raggedright{w/o Spatial Descriptors +w/o Matching Loss}
&  \centering{0.555} & \centering{{0.096}} & \centering{0.424} & \cellcolor{LightGray}\centering{0.359} 
& \centering{0.829} & \centering{{0.127}} & \centering{0.666} & \cellcolor{LightGray}\centering{0.541} 
\cr
w/o Matching Loss  
&  \centering{0.592} & \centering{0.126} & \centering{{0.402}} & \cellcolor{LightGray}\centering{0.373} 
& \centering{0.861} & \centering{0.176} & \centering{0.683} & \cellcolor{LightGray}\centering{0.573} 
\cr
w/o K-function Descriptor 
&  \centering{0.417} & \centering{0.154} & \centering{0.431} & \cellcolor{LightGray}\centering{0.334} 
& \centering{{0.602}} & \centering{0.226} & \centering{0.583} & \cellcolor{LightGray}\centering{0.470} 
\cr
Ours 
&  \centering{{0.413}} & \centering{0.146} & \centering{{0.357}} & \cellcolor{LightGray}\centering{\textbf{0.306}} 
& \centering{0.611} & \centering{0.201} & \centering{{0.509}} & \cellcolor{LightGray}\centering{\textbf{0.440}} 
\cr
\hline
\end{tabular}
\caption{Evaluation of Cross K-function of generated cell layout compared to the reference layouts in BRCA-M2C dataset.}
\label{table:layout-eval-K}
\end{center}
\end{table*}

\subsection{Implementation Details}
\myparagraph{Layout generator.}
Our cell layout generator model is based on the point cloud generator SP-GAN~\cite{li2021spgan}. We make several changes to the model to make it suitable for the conditional cell layout generation task. SP-GAN takes as input a fixed 3D point cloud in the form of a unit sphere. Instead, We have varying size 2D points with their pre-assigned classes. The coordinates for the points in each class are equally distributed in a mesh grid in the range $(-1,1)$ with a small normal perturbation. The mesh size varies for each class based on the number of points in the class so the points end up covering the space. The conditioning spatial descriptors are transformed to a 32 dimensional vector embedding before being attached to every point. Last, to account for the variation in input sizes, we employ instance norm instead of batch norm. We use 2 discriminators; $D$ and $D_c$ for adversarial training.  $D$ discriminates a layout or a set of coordinates as real/fake and is ignorant of the points classes. $D_c$ is similar but takes also the points classes into account. Using least squares GAN loss, the Generator loss is a weighted sum of the GAN losses and the cell configuration loss. 
\begin{equation}
\begin{split}
    \mathcal{L_G}=&\frac{1}{2}[(D(\hat{P})-1)^2 + (D_c(\hat{P}_{c})-1)^2\\ &+(L_{CC}(\hat{P}_{c},P_{c}))]    
\end{split}
\end{equation}
where $P$ and $\hat{P}$ are the reference real, and generated point sets, respectively. $P_{c} and \hat{P}_{c}$ are the reference real, and generated point sets along with the class assigned to each point.

\myparagraph{Layout to Image Generator.}
The trained cells' layout generator provides realistic cell layouts that can be used by themselves as augmentations to train a machine learning model on multi-class point patterns. However, to use the generated layouts to provide more complex image augmentation than flipping, rotation, or stain/style transfer, we need to transform the layouts into images. To do that, we create a layout-to-image generator model based on the pix2pix model~\cite{pix2pix2017}, and its variation in the biomedical domain~\cite{chang2020synthetic}. There are 3 main differences from \cite{pix2pix2017} in our setting, first we do not have an exact mask of the cell shapes, we only have their coordinates. Second, while ~\cite{pix2pix2017} learns a specific mapping from one domain to the other, here the generated images are expected to have a similar texture as a reference input H\&E image, and last, there are relatively few annotated images for training with biomedical data.

To generate an image from the cells coordinates, we first create  binary images; one per class with a point at each cell location and dilate these points to give us a visible layout. These multi-channel cell layout together with a reference H\&E image are the input to the Layout-to-Image generator. The output is an image with the similar texture as the reference H\&E image and has the same cells distribution as the layout image. The model is trained adversarially with multi-scale discriminators that classify whether an image is real or fake and whether $2$ images come from the same slide. For annotated images, we use L1  reconstruction loss, in addition to a perceptual loss, similar to~\cite{chang2020synthetic}.

\subsection{Dataset}
We use the breast cancer dataset, BRCA-M2C~\cite{Abousamra_2021_ICCV}. It 
consists of 120 patches belonging to 113 patients,  collected from TCGA~\cite{tcga}. The patches are $\approx 500\times500$ pixels at $20$x magnification, which is large enough to
provide spatial context. The annotations 
are in the form of dot annotations at the 
approximate centers of cells. Each dot is assigned one of $3$ main cell classes: inflammatory, epithelial, or stromal. 

\subsection{Evaluation of Cell Layout Generation}
To evaluate the quality of the generated cell layouts, we propose a set of metrics to measure the similarity of the spatial distributions of the generated and the reference layouts. We focus on both topology and spatial statistics.

For topology, we compare a generated layout and its reference layout by comparing their persistence diagrams. Our evaluation is carried out class-by-class. We compare diagrams for each class of cells, and aggregate the scores.
To compare two diagrams, we use two metrics: Earth Mover's Distance (PD-EMD) and Cell Configuration Matching Distance (PD-CCMD). PD-EMD is agnostic to the multi-class spatial configuration surrounding each hole and so it may not give an accurate evaluation. To get a better evaluation, we propose to use the cell configuration matching distance in PD-CCMD, which is designed to take into account the spatial configuration. It matches the holes in the generated and reference layouts using the optimal K-function matching $\gamma^\ast$ as in Eq.~\eqref{eq:opt-matching}. Next, PD-CCMD computes the mean distance between the persistence of the matched holes. 
Note that the persistence  distance is ignored when the number of cells in the class is very small (less than 5), and is assumed to be zero. This is because when there are very few points, the distance is unreliable and greatly affected by where the points are located with respect to the border and hence can result in very large unreasonable distance values.
Table~\ref{table:layout-eval-PD} shows the PD-EMD and PD-CCMD metrics for each class of cells and their mean.

To evaluate the spatial co-localization across different classes, we use the cross K-function. For each class of cells, cross K-functions are computed with that class as the source and  different target classes, creating a high dimensional vector. The mean absolute error (MAE) and root mean squared error (RMSE) are computed between the generated and real layouts vectors, as shown in Table~\ref{table:layout-eval-K}. 
The distance is normalized by the vector dimensions and the expected number of cells from one class in a patch.

We evaluate the generated layouts using our proposed metrics in Table~\ref{table:layout-eval-PD} and Table~\ref{table:layout-eval-K}. We compare our proposed method to models trained: (a) without spatial descriptors and without the cell configuration loss, (b) with spatial descriptors but without cell configuration loss, (c) with the cell configuration loss but without the K-function spatial descriptor. Adding the cell configuration loss improves performance, and the best result uses both the K-functions and the multi-scale densities. We also observe that without the cell configuration loss, the model tends to collapse, generating almost identical layouts as the reference layouts.  

\subsection{Cell Layout Generation for Augmentation}
We test the generated cell layouts on the downstream cell classification task on the BRCA-M2C dataset. We use the multi-class cell layouts generated by our cell layouts generator and use the layout-to-image generator to transform into H\&E images that can be used for data augmentation. 
To train the layout-to-image generator, we use the labeled patches in BRCA-M2C in the reconstruction and perceptual losses. We extract additional patches from TCGA; some from nearby the annotation regions in the same slides and others randomly sampled from different slides, and ensure that sampled patches do not belong to the background. During training, the reference texture patch and the reference cell layout patch may belong to the same slide or may come from different slides. The perceptual and reconstruction losses are only applied when they are from the same slide, while the adversarial losses are applied in both cases. To generate the H\&E images for augmentation, we generate cell layouts and apply a postprocessing to remove overlapping cells. The final layout along with varying reference H\&E patches are transfomred into H\&E patches with different styles, see Figure~\ref{fig:results-texture}.


We train U-Net~\cite{unet:2015:miccai} and MCSpatNet~\cite{Abousamra_2021_ICCV} on the BRCA-M2C dataset in addition to augmentation data generated from our models and from random cell layouts. The loss is weighted based on whether it is real or generated data, giving the generated data a lower weight of 0.5. Table~\ref{table:BRCA-M2C-results} shows the F-score comparing both models trained with and without our data augmentation. We see that the augmentation improves the performance, especially with the U-Net model, with greater improvement when using our generated augmentation data. For MCSpatNet, the Stromal cells F-score is below the F-score without the augmentation. We attribute this to the quality of the image generation. The image generation model needs to learn how each type of cells appear and that is a challenging task specially with Stromal cells. They are often hard to classify without uncertainty even by expert pathologists.

\begin{table}[h!]
\small
\begin{center}
\begin{tabular}{|p{0.42\linewidth}|p{0.09\linewidth}|p{0.09\linewidth}|p{0.09\linewidth}|p{0.09\linewidth}|
}
\hline
 Method  & \centering{Infl.} & \centering{Epi.} & \centering{Stro.} & \cellcolor{LightGray}\centering{Mean} 
 \cr
\hline
U-Net 
& \centering{0.498} & \centering{0.744} & \centering{0.476} & \cellcolor{LightGray}\centering{0.572} 
\cr
U-Net + Aug. (Rand.) 
& \centering{0.625} & \centering{0.735} & \centering{0.472} & \cellcolor{LightGray}\centering{0.611} 
\cr
U-Net + Aug. (Ours) 
& \centering{\textbf{0.65}} & \centering{\textbf{0.768}} & \centering{\textbf{0.511}} & \cellcolor{LightGray}\centering{\textbf{0.644}} 
\cr
\hline
MCSpatNet  
& \centering{{0.635}} & \centering{{0.785}} & \centering{\textbf{0.553}} & \cellcolor{LightGray}\centering{{0.658}} 
\cr
MCSpatNet + Aug. (Rand.) 
& \centering{\textbf{0.652}} & \centering{{0.772}} & \centering{0.506} & \cellcolor{LightGray}\centering{{0.644}} 
\cr
MCSpatNet + Aug. (Ours) 
& \centering{\textbf{0.678}} & \centering{\textbf{0.8}} & \centering{0.522} & \cellcolor{LightGray}\centering{\textbf{0.667}} 
\cr
\hline
\end{tabular}
\caption{F-scores on the cell classification task, comparing models trained with only manually labeled data to models trained with additional data  augmentation from random cell layout (Rand.) and from our generated cell layout (Ours).}
\label{table:BRCA-M2C-results}
\end{center}
\vspace{-.25in}
\end{table}

\section{Conclusion}
\label{sec:conclusion}
In this paper, we propose the first generative model for digital pathology to explicitly generate cell layout with desirable configuration. We focus on topological pattern and spatial distribution of multi-class cells, and compute configuration descriptors based on classic spatial statistics and theory of persistent homology. Using these descriptors, and by proposing a novel cell configuration loss, our generator can effectively generate new cell layouts based on a reference cell layout. We show through qualitative and quantitative results that our method generates cell layouts with realistic spatial and structural distribution. We also use our method to augment H\&E images, thus improving the performance in downstream tasks such as cell classification.

\myparagraph{Acknowledgement.} This work was support by the NSF grants CCF-2144901, IIS-2123920 and IIS-2212046, the National Institutes of Health (NIH) and National Cancer Institute (NCI) grants UH3CA225021, U24CA215109, 5R01CA253368, as well as generous private fund from Bob Beals and Betsy Barton.

{\small
\bibliographystyle{ieee_fullname}
\bibliography{egbib,refs}
}

\end{document}